\documentstyle[pra,aps,preprint]{revtex}
\begin{document}
\draft
\title{Lattice Instability in the Spin-Ladder System under Magnetic Field}
\author{Naoto Nagaosa and Shuichi Murakami}
\address{Department of Applied Physics, University of Tokyo,
Bunkyo-ku, Tokyo 113, Japan}
\date{\today}
\maketitle
\begin{abstract}
We study theoretically the lattice instability in the spin gap systems under
magnetic field. With the magnetic field larger than a critical value 
$h_{c1}$, the spin gap is collapsed and the magnetization arises.
We found that the lattice distortion occurs in the spin-ladder 
at an incommensurate wavevector corresponding to the magnetization, 
while it does not occur in the Haldane system.
At low temperatures the magnetization curve 
shows a first order phase transition with this lattice 
distortion.
\end{abstract}

\narrowtext
The one-dimensional (1D) quantum spin systems with the gap
in the excitation spectrum have been attracting intensive interest
both theoretically and experimentally. This is due to the underlying large 
quantum fluctuations which destroy the magnetic long range order (LRO)
\cite{hal}. 
Another aspect of these spin gap systems is that they are free from the 
instabilities, e.g., spin-Peierls instability \cite{cross}, 
which comes from the diverging 
generalized susceptibility as the temperature $T \to 0$.
One of the spin gap systems of the current interest is the 
antiferromagnetically coupled two-leg spin ladder system 
\cite{dag,takano,ham1,cha1,cha2}. 
Under a magnetic field at $T=0$ the magnetization 
remains zero until $h$ reaches a critical value $h_{c1}$ because the 
ground state is the spin singlet.
At $h=h_{c1}$ the singlet and lowest triplet states are degenerate, and the 
magnetization begins to appear. Between $h_{c1}$ and $h_{c2}$, where  the 
magnetization increases,
the system remains gapless and is described
as a Tomonaga-Luttinger liquid with the continuously changing exponent. 
For $h>h_{c2}$ the system is perfectly spin polarized.
Recently the two-leg ladder spin system Cu$_2$(1,4-Diazacycloheptane)$_2$Cl$_4$
has been studied under magnetic field \cite{cha1,crowell,ham2,hagiwara,chiba}.
The magnetization process  
has been studied also theoretically \cite{weihong,hayward,gia}.
In addition, an anomaly of  the specific heat is observed at $T=1K$ when 
$h = 9.0T> h_{c1} = 7.2T$ \cite{ham2}.
This strongly suggests the phase transition to the ordered state, 
which is assumed to be AF long range ordering \cite{cha1}, but
is not yet confirmed.

Motivated by these experimental works, we study in this letter 
the alternative instability, i.e., lattice instability,  of the two-leg spin 
ladder system in its gapless region $h_{c1}<h<h_{c2}$. 
We found that when coupled to the 
lattice distortion, which should be always there, the gapless state is unstable
to the (generalized) spin-Peierls state accompanied with the modulation of the 
inter-chain exchange interaction with the  
incommensurate wavenumber proportional to the magnetization.
Another consequence of this instability is that the magnetization curve
at low temperature shows a first order phase transition, and the 
schematic phase diagram is given in Fig. 1.
This is in sharp contrast to the Haldane system, where there is no such an
instability.   

We start with the bosonized Hamiltonian for the two coupled 
spin 1/2 Heisenberg model \cite{strong,wata,naga1,tsv}.
\begin{equation}
H = H_+ + H_- + H_{\rm lattice}
\end{equation}
\begin{equation}
H_+ = \int d x \biggl[  { {v_s} \over 2 } ( \Pi_+^2 + (\partial_x \phi_+)^2 )
- { m \over {\pi a_0}}  \cos \sqrt{ 4 \pi } \phi_+
- { {h \partial_x \phi_+} \over { \sqrt{\pi}} } \biggr]  
\end{equation}
\begin{eqnarray}
H_- &=& \int d x \biggl[ 
{ {v_s} \over 2 } ( \Pi_-^2 + (\partial_x \phi_-)^2 )
+ { m \over {\pi a_0}} \cos \sqrt{ 4 \pi } \phi_-
\nonumber \\
&+& { {2m}  \over {\pi a_0}} \cos \sqrt{ 4 \pi } \theta_- \biggr]
\end{eqnarray}
where $\phi_j$ and $\theta_j$ are canonical conjugate phase fields
for $j$-th chain, and
$\phi_{\pm} = (\phi_1 \pm \phi_2)/\sqrt{2}$,
$\theta_{\pm} = (\theta_1 \pm \theta_2)/\sqrt{2}$.
$v_s$ is the spin wave velocity, and $a_0$ is the lattice constant which we 
put to be 1. 
The mass $m$ represents the spin gap, which originates from the 
inter-chain exchange interaction $J_{\perp}$, i.e.,
\begin{equation}
m \cong  { {J_{\perp}} \over {2 \pi}}
\end{equation}
Antiferromagnetic $J_{\perp}$ corresponds to the two-leg spin ladder, while
the (strong) ferromagnetic $J_{\perp}$ to the Haldane system.
The magnetization density $M(x)$, which is coupled to the 
magnetic field $h$,  is given by the symmetric part
as $M(x) = \partial_x \phi_+(x)/ \sqrt{\pi}$.  
It is known that both $H_+$ and $H_-$ have the massive
spectra when $h=0$. Only $H_+$ is modified by the magnetic field $h$
and becomes gapless, while the antisymmetric part $H_-$ remains unchanged.
The coupling to the lattice distortion is given by
\begin{eqnarray}
H_{\rm lattice} &=& \int d x \sum_{j=1,2} \biggl[
{ 1 \over 2} u_j(x)^2
- { g \over {\pi a_0} }  u_j(x) \sin \sqrt{2 \pi} \phi_j(x) \biggr]
\nonumber \\
&+&  \int d x \biggl[ { 1 \over 2} v(x)^2 - 
{{\gamma} \over { \pi a_0} } v(x) 
[- \cos \sqrt{4 \pi} \phi_+(x) 
\nonumber \\
&+& \cos \sqrt{ 4 \pi} \phi_-(x) + 
2 \cos \sqrt{4 \pi} \theta_-(x) ] \biggr]
\end{eqnarray}
where $u_j$ represents the dimerization of the $j$-th chain which modulates 
the intrachain exchange interaction, while $v$ the 
change of the interchain distance which  modulates the 
interchain exchange interaction.
We treat the fields $u_j$, $v$ classically, which is justified because the 
lattice is usually three-dimensional.
It can be seen from eq.(2) that the magnetic field will introduce solitons
when the magnetic energy gain is larger than the soliton formation energy.
To make it more explicit we transform $H_+$ and the coupling to the 
interchain modulation $v$ into the Fermionic form by
introducing the Fermion operator 
$
\psi_{R,L}(x) = ( 2 \pi a_0)^{-1/2} \exp [ \pm i \sqrt{\pi} 
 (\phi_+\mp \theta_+ ) ]
$ \cite{tsv}:
\begin{eqnarray}
H_+^{\rm Fermion} &=& \int d x 
\biggl[ - i v_s 
( \psi_R^\dagger \partial_x \psi_R - \psi_L^\dagger \partial_x \psi_L)
\nonumber \\
&-& i ( m + \gamma v(x) )( \psi_R^\dagger  \psi_L - \psi_L^\dagger \psi_R)
+ { 1 \over 2} v(x)^2
\nonumber \\
&-& h ( \psi_R^\dagger \psi_R + \psi_L^\dagger \psi_L) \biggr]
\end{eqnarray}
This transformation is possible because $H_+$ describes the 
sine-Gordon model at $\beta^2 = 4 \pi$ \cite{tsv}.
Equation (6) represents the Dirac Fermion with the mass $m$ 
coupled  to the lattice distortion $v(x)$ with the Fermi energy controlled 
by the magentic field $h$.  
When $h$ increases beyond $h_{c1} = m$, 
the Fermi level lies in the upper band of the Dirac Fermion. 
The magnetization $M$ is represented by the Fermi wavenumber $k_F$ as
$k_F = \pi M$ where $M$ is the normalized to be unity when all the 
spins are polarized.
Then the generalized spin-Peierls transition occurs 
at the nesting wavevector $Q = 2 k_F = 2 \pi M$, i.e.,
\begin{equation} 
v(x) = 2 v_0 \cos (2 \pi M x + \delta)
\end{equation}
with $\delta$ being some phase. The displacement is schematically 
shown in Fig. 2.
First consider the case where the magnetization $M$ 
is not too small and the 
Fermi energy $\varepsilon_F$ is large enough compared with the 
transition temperature $T_c$.
A mean field treatment results in a second
order phase transiton at $T = T_c \cong \varepsilon_F
e^{ - 1/(N(\varepsilon_F)\gamma^2) }$ with the specific heat jump
$\Delta C/ C(T_c) = 1.42$ where $C(T)$ is proportional to $T$.
The validity of this weak coupling mean field 
theory can be examined as follows. 
The experiment has been done for
$h = 9.0T$ where the magnetization $M \cong 0.3$ \cite{cha1,ham2,hagiwara}. 
The gap $m \cong 0.89 meV$ and the spin wave velocity
$v_s$ is estimated from the intrachain exchange interaction $J_2 = 0.21 meV$
as $v_s = { \pi \over 2} J_2 \cong 0.33 meV$ \cite{ham2}.
Therefore $v_s k_F = \pi v_s M \cong 0.3 meV$, which is about 
one-third of $m$. The Fermi energy $\varepsilon_F$ measured from the 
bottom of the upper band is then $\varepsilon_F \cong 0.05 meV \cong 0.5K$.
This estimate shows that the transition temperature $T_c$ is the 
same order as the Fermi energy $\varepsilon_F$, and 
the system is in the intermediate coupling regime.

As the magnetic field is decreased to $h_{c1}$ the 
Fermi energy $\varepsilon_F$ also decreases and the 
mean field treatment breaks down. 
Then we consider the zero temperature limit 
with changing the magnetic field.
When the magnetization $M$ is small and 
the off-diagonal matrix element $\gamma v_0$ introduced by the distortion 
is larger than the Fermi energy $\varepsilon_F$, many $k$ points are 
connected by that matrix element.
This is due to the parabolic dispersion which becomes important
near the bottom of the band.
As shown below the kinetic energy gain 
due to the distortion is given by $4 \pi M \gamma v_0$
and $v_0$ is determined as $v_0 = 2 \pi \gamma M$.
With this $v_0$, the matrix element is 
$2 \pi \gamma^2 M$ and equating this to the dispersion
$ (v_s k_0)^2/2m$ we obtain $k_0 \cong \sqrt{4 \pi m \gamma^2 M}/v_s$
The number $G$ of the reduced Brillouin zones within 
the region $[-k_0, k_0]$ is
$G \cong  k_0/(\pi M) \cong \sqrt{4  m \gamma^2/(\pi v_s^2 M)}$
and $G>>1$ in the limit of small $M$.
In this case we can neglect the dispersion $(v_s k)^2/2m$ in the
diagonal matrix element compared with the off-diagonal ones, and 
the eigen-value problem can be approximately regarded as 
that for the tight binding wave with the wavenumber $q$ discretized as
$q_n = 2 \pi n/ G$ ($n$: integer). Then the eigen-value is given by 
\begin{equation}
\varepsilon_n = - 2 v_0 \gamma \cos q_n.
\end{equation}
Therefore the lowest band ($q = n = 0$) with the energy $- 2\gamma v_0$ is 
occupied and the energy gain is $- 4 \pi \gamma v_0 M$.
Then minimizing $ v_0^2 - 4 \pi \gamma v_0 M$, we obtain
$v_0 = 2\pi \gamma M$ as mentioned above.
The gap between the lowest ( $n=0$) and the second lowest one ($n=1$)
is of the order of $ v_0 \gamma / G^2 \propto M^2$
and is much smaller than the matrix element $v_0 \gamma (\propto M)$
in the limit of small $M$. 
This happens when the magnetic field $h$ is slightly above the 
critical field $h^{(0)}_{c1}$ without the lattice distortion  
and we define $\Delta h = h - h^{(0)}_{c1}$. 
Based on the above discussion, let us  consider $\Delta h$ as a function of
the magnetization $M$. Without the lattice distortion, 
$\Delta h = (\pi v_s M)^2/( 2 m)$ for small $M$.
With the coupling to the lattice, it changes into 
\begin{equation}
\Delta h = - 2 \gamma v_0 + \alpha (\pi v_s M)^2/ m
=- 4 \pi \gamma^2 M + \alpha (\pi v_s M)^2/ m 
\end{equation}
where $\alpha $ is a constant of the order of unity.
This is because the chemical potential at $T=0K$ is at the middle of the gap
between the lowest ($n=0$) and the second lowest $(n=1)$ bands
which is 
$ - 2 \gamma v_0 + O( (v_s M)^2/m)$.
The relation (9) means that the 
magnetization $M$ jumps from 0 to a finite value 
$M = M_0 \sim  m \gamma^2/ v_s^2$ with a first order phase transition.
Summarizing the considerations above, the schematic phase diagram for 
spin ladder system coupled to the lattice is given in Fig. 1.
The first order phase transition line terminates at a tricritical point
to change into the second order transition line.

There is no instability for the dimerization $u_j$, 
which corresponds to the standard spin-Peierls distortion, when the   
coupling constant $g$ is small enough. This is because the 
operator $\sin \sqrt{2 \pi} \phi_j = \sin \sqrt { \pi} ( \phi_+ \pm \phi_-)$
contains $\cos \sqrt{\pi}\phi_-$ and $\sin \sqrt{\pi} \phi_-$. 
In ref.\cite{tsv} it is discussed via the mapping 
the Ising model that the correlation functions for 
$\cos \sqrt{\pi} \phi_-$ and 
$\sin \sqrt{\pi} \phi_-$ decay exponentially.
Then also the correlation
function of $\sin \sqrt{2 \pi} \phi_j$ decays exponentially.
This means that the energy gain 
due to the dimerization $u_j$ is given by $\sim g^2 u_j^2$ for small $g$
even at zero temperature, and the system is stable against the 
dimerization as long as $g$ is small because of the elastic energy $u_j^2/2$.
Therefore, the lattice instabillity does not occur in the Haldane system
where the interchain coupling $J_{\perp}$ represents the 
ferromagnetic Hund's coupling and is very strong ($\sim 1 eV$).
The spin gap is due to the quantum fluctuation of the resulting $S=1$ spins
\cite{hal}, and the modulation of $J_{\perp}$ does not play any role. 
Then the magnetization process can be discussed by a model without the 
coupling to the lattice \cite{goto,sakai}.
Another instability is of course the AF ordering as discussed in ref.
\cite{cha1}.
The competition between the lattice instability and the 
AF is determined by the relative magnitude of the inter-ladder  
exchange interactions and the  spin-lattice coupling, and the 
coexistence of these two is not expected for clean system \cite{ina}. 
     
We now discuss the experimental consequences of the above scenario.
(i) The Bragg spot should be observed in the X-ray and/or neutron 
scattering at the wavevector $Q = 2 \pi M$
due to the modulation of the interchain distance.
(ii) There appears a gap below $T_c$ and the specific heat should behave 
as $C \sim e^{ - \gamma v_0/ T}$, where $\gamma v_0$ is the 
gap of the order of the transition temperature $T_c$.
(iii) The magnetization curve  shows a hysteresis behavior at low temperatures.
The experimental confirmation of these predictions is highly desired.

In conclusion we have studied the lattice instability in the
spin-ladder system when the spin gap is collapsed by the magnetic field.
The lattice will distort to modulate the interchain exchange interaction with 
the incommensurate wavenumber $Q = 2 \pi M$.

The authors acknowledge M.Hagiwara, H.Fukuyama, Y. Tokura,
M. Yamanaka  for fruitful discussions. 
This work is supported by Grant-in-Aid for 
Scientific Research No. 05044037, No. 04240103, and No. 04231105
from the Ministry of Education, Science, and Culture of Japan.

\vfill
\eject
\noindent
Figure captions
\par 
\medskip
\noindent
Fig. 1: 
 Schematic phase diagram of the two-leg spin ladder system 
coupled to the lattice distortion in the plane of temperature $T$ and the 
magnetic field $h$. The thick solid curve is the first order phase transition 
line while the broken one the second order transition between the 
undistorted and distorted states. 
The magnetization jumps across the first order transition line. 
\par
\par
\medskip
\noindent
Fig. 2: 
Schematic view of the lattice distortion which modulates the 
interchain exchange interaction. The wavenumber $Q = 2 \pi M$
changes as the magnetization $M$ increases.
\par


\begin{thebibliography}{99}

\bibitem{hal}F. D. M. Haldane, Phys. Rev. Lett {\bf{50}} (1983) 1153.
\bibitem{cross}M. C. Cross and D. S. Fisher, Phys. Rev. B{\bf{19}} (1979) 402.
\bibitem{dag}E. Dagotto, J.Riera and D.J.Scalapino, 
Phys. Rev. B{\bf{45}} (1992) 5744.
\bibitem{takano}M. Takano et al., JJAP Series {\bf{7}} (1992) 3.
\bibitem{ham1}P. R. Hammar and D. H. Reich, J. Appl. Phys. 
{\bf{79}} (1996) 5392.
\bibitem{cha1}G. Chaboussant et al., Phys. Rev. B{\bf{55}} (1997) 3046.
\bibitem{cha2}G. Chaboussant et al., Phys. Rev. Lett. {\bf{79}} (1997) 925.
\bibitem{crowell}P. A. Crowell et al., Rev. Sci. Instrum. {\bf{67}} 
(1996) 4161.
\bibitem{ham2}P. R. Hammar, D. H. Reich, C. Broholm, and F.Trouw, 
cond-mat/9708053.
\bibitem{hagiwara}M. Hagiwara, Y. Narumi, K. Kindo, T. Nishina, M. Kaburagi, 
and T. Tonegawa, unpublished.
\bibitem{chiba}M. Chiba, T. Fukui, Y. Ajiro, M. Hagiwara, T. Goto, and 
T. Kubo, to appear in Proc. 5th Symposium on Research in High Mag. Field
( Sydney Aug, 1997); M. Chiba, T. Kubo, M. Hagiwara, Y. Ajiro, T. Asano, 
and T. Fukui, to appear in Proc. Int. Conf. on Magnetism ( Cairns, July 1997).
\bibitem{weihong}Z. Weihong, R. R. P. Singh, J. Oitmaa, Phys. Rev. 
B{\bf{55}} (1997) 8052.
\bibitem{hayward}C. A. Hayward, D. Poilblanc, and L. P. Levy, 
cond-mat/9606145.
\bibitem{gia}R. Chitra and T. Giamarchi, Phys. Rev. B{\bf{55}} (1997) 5816.
\bibitem{strong}S. P. Strong and A. J. Millis, 
Phys. Rev. Lett. {\bf{69}} (1992) 2419.
\bibitem{wata}H. Watanabe, K. Nomura and S. Takada, 
J. Phys. Soc. Jpn. {\bf{62}} (1993) 2845.
\bibitem{naga1}N. Nagaosa, Solid State Commun. {\bf{94}} (1995) 495; 
N. Nagaosa and M. Oshikawa,
J. Phys. Soc. Jpn. {\bf{65}} (1996) 2241.
\bibitem{tsv}D. G. Shelton, A. A. Nersesyan, and A. M. Tsvelik, 
Phys. Rev. B{\bf{53}} (1996) 8521.
\bibitem{goto}T. Goto, H. A. Katori, and Y. Ajiro, J. Phys. Soc. Jpn. 
{\bf{61}} (1992) 4155.
\bibitem{sakai}T. Sakai and M. Takahashi, 
Phys. Rev. B{\bf{43}} (1991) 13383.
\bibitem{ina}S. Inagaki and H. Fukuyama, N. Nagaosa, J. Phys. Soc. Jpn. 
{\bf{52}} (1983) 3620. 

\end{thebibliography}
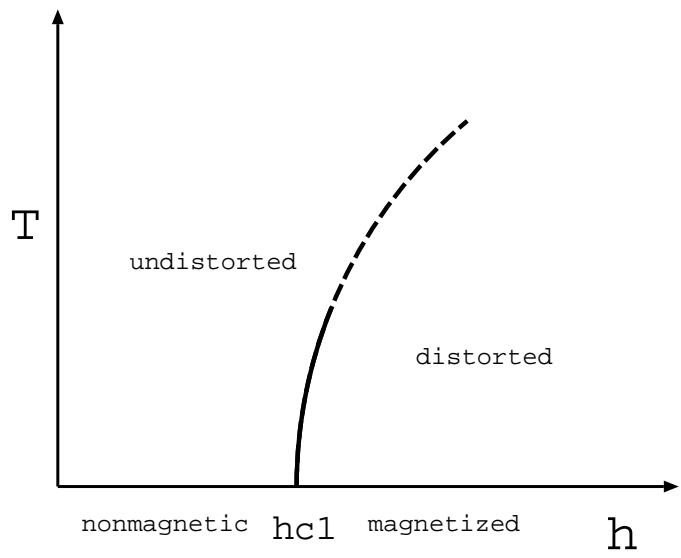
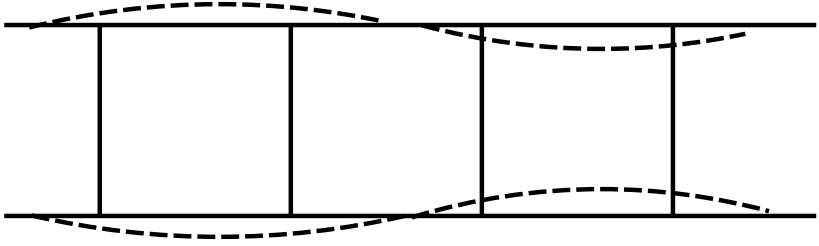
\end{document}